\documentclass[11pt]{article}

\makeatletter
\providecommand{\@LN}[2]{}
\makeatother

\usepackage[a4paper,hmargin=2.5cm, vmargin=2.5cm]{geometry}
\usepackage[dvipsnames]{xcolor}
\usepackage[colorlinks,linkcolor={blue},citecolor={blue}]{hyperref}
\usepackage{amssymb,amsmath,amsthm}
\usepackage[mathscr]{eucal}
\usepackage{float}
\usepackage{graphicx}
\usepackage{booktabs}
\usepackage{natbib}
\usepackage{enumitem}
\usepackage{multirow}

\newcommand{\diff}{\mathrm{d}}

\newcommand{\T}{\mathrm{\scriptscriptstyle T}}

\newcommand{\bfxy}[1]{\mathbf{#1}}
\newcommand{\bfgk}[1]{\pmb{#1}}

\newcommand{\bfX}{\bfxy{X}}
\newcommand{\bfx}{\bfxy{x}}

\newcommand{\bfu}{\bfxy{u}}

\newcommand{\bfp}{\bfxy{p}}

\newcommand{\E}{\mathbb{E}}
\newcommand{\pr}{\mathbb{P}}

\newcommand{\cond}{\,|\,}

\DeclareMathOperator*{\argmax}{\mathrm{argmax}}

\newtheorem{theorem}{Theorem}

\theoremstyle{definition}

\title{
    Integration of Individual Participant and Aggregate Data Under Dataset Shift: Summary Statistic Comparison and Scalable Computation
}
\author{
    Ming-Yueh Huang\\
    {\it Institute of Statistical Science, Academia Sinica, Taipei 11529, Taiwan}\\[10pt]
    Jing Qin\\
    {\it Biostatistics Research Branch, National Institute of Allergy and Infectious Diseases,}\\
    {\it National Institutes of Health, Bethesda, Maryland 20892, U.S.A.}\\[10pt]
    Chiung-Yu Huang\\
    {\it Department of Epidemiology and Biostatistics,}\\
    {\it University of California, San Francisco, CA 94158, U.S.A.}
}

\date{}

\begin{document}

\maketitle

\begin{abstract}
    Integrated IPD–AD analysis, which combines individual participant data (IPD) with aggregate data (AD), is increasingly recognized as an effective strategy for generating more reliable and generalizable inferences from heterogeneous studies. 
    While most existing work has focused on algorithmic approaches, this paper investigates a complementary yet underexplored question: how different forms of AD influence the efficiency of data integration.
    Working within a constrained maximum likelihood estimation framework, we compare commonly reported summary statistics and show that subgroup-specific summaries can substantially improve estimation efficiency. 
    In particular, we find that AD derived from outcome-stratified subgroups (e.g., cases and controls) consistently yield greater efficiency gains than those based on covariate-stratified subgroups (e.g., age or exposure categories), especially when the outcome is continuous.
    Although outcome-stratified summaries are commonly reported for discrete outcomes, they are rarely provided when the outcome is continuous.
    Our findings therefore support the routine inclusion of outcome-stratified summaries for continuous endpoints in trial reports and public data repositories to facilitate more efficient evidence synthesis.
    We further extend the constrained maximum likelihood framework to accommodate dataset shift and develop a fast, non-iterative estimation procedure to improve numerical stability and scalability.
    We illustrate the proposed methodology with two applications: an analysis of income data under covariate shift and an analysis of housing data under prior probability shift.
\end{abstract}

\section{Introduction}

Integrating individual participant data (IPD) with aggregate data (AD) has emerged as an important strategy for synthesizing evidence across diverse studies.
While IPD provides detailed participant-level information that enables flexible subgroup analyses and better control of potential confounding, access to such data is often limited due to privacy, regulatory, or logistical constraints \citep{AhmedSR2012Assessment}.
In contrast, AD consists of study-level summary statistics that are typically easy to obtain but lack the granularity needed for detailed investigation \citep{SuttonH2008Recent}.
An integrated IPD–AD analysis maximizes data utilization by combining detailed information from IPD with study-level AD in a single modeling framework, allowing researchers to capitalize on the strengths of both data types to improve the efficiency and generalizability of statistical inference.

A central methodological challenge in IPD–AD integration is that AD can be reported in different forms, ranging from simple descriptive statistics to model-based estimates, each subject to sampling variability and each carrying different information about the underlying parameters.
Constrained maximum likelihood estimation (CMLE) provides a general framework for combining such heterogeneous summaries with IPD \citep{Qin2000Combining, ChatterjeeCMC2016Constrained}.
In this framework, AD is viewed as estimates of parameters that satisfy unbiased population estimating equations, which are then imposed as constraints on the likelihood based on IPD to improve efficiency.
Uncertainty in AD can be handled by assuming the summary estimates are approximately normal and including the corresponding likelihood contribution in the constrained objective \citep{ZhangDSQY2020Generalized}.

Another key challenge in integrated IPD–AD analysis is heterogeneity across datasets, especially when the IPD and AD are collected from different sources, study populations, or time periods \citep{ChenNSQ2021Combining}.
Although between-study heterogeneity is widely acknowledged in the meta-analysis literature and is often modeled through random-effects frameworks, systematic evaluation of specific forms of heterogeneity has been limited.
In this paper, we conceptualize between-study heterogeneity using the dataset-shift framework from machine learning \citep{Storkey2008When,Finlaysonetc2021Clinician} and consider a general dataset shift model that includes several important types of shift, such as covariate shift \citep{ShengSHK2022Synthesizing, ChengLTH2023Semiparametric} and prior probability shift \citep{ChenHCSQ2024Integrating}.
Within this framework, we use a general density ratio model \citep{QinZ1997Goodness,Qin1998Inferences} to link the data-generating mechanisms for IPD and AD, derive unbiased estimating equations for AD, and develop unified CMLE procedures that yield valid inference under these shifts.

While much of the existing literature on IPD–AD meta-analysis has focused on developing integration algorithms, less attention has been paid to how different forms of AD affect the efficiency of statistical inference.
To address this gap, we compare three commonly used types of summaries within a unified CMLE framework: marginal means of outcomes or covariates \citep{ImbensL1994Combining,Qin2000Combining}, covariate-stratified outcome summaries \citep{HuangQT2016Efficient,ShengSHK2022Synthesizing,ChengLTH2023Semiparametric}, and outcome-stratified covariate summaries.
We evaluate their performance both in the absence of dataset shift and under the general shift model, and show that outcome-stratified covariate summaries can substantially improve estimation efficiency relative to the other two classes.
Although such summaries are routinely reported in case–control studies (i.e., for binary outcomes), they are much less commonly provided for continuous outcomes.
Our results therefore provide strong support for routinely reporting outcome-stratified summaries for continuous endpoints in trial publications and public data repositories to facilitate more efficient evidence synthesis.

Intuitively, the potential efficiency gain increases as more constraints are incorporated into the CMLE formulation.
However, when rich AD is used, for example outcome-stratified covariate summaries with many covariates and finely partitioned outcomes, the number of constraints can become very large, leading to high-dimensional Lagrange multiplier systems and substantial numerical instability.
As a result, standard CMLE algorithms, although flexible and theoretically appealing for integrating IPD and AD, can be computationally demanding and difficult to scale in realistic applications.
To address this challenge, we build on the non-iterative strategy of \cite{GaoC2023Noniterative} and develop a fast algorithm that obtains CMLE in a single step while accommodating both AD uncertainty and dataset shift simultaneously.
The resulting procedure retains the theoretical advantages of CMLE and substantially improves numerical stability and scalability in high-dimensional integration problems.

The rest of this article is organized as follows.
Section \ref{sec:CMLE} summarizes the approach for leveraging AD while accounting for uncertainty in the data.
In Section \ref{sec:DatasetShift}, we further extend the CMLE under covariate shift and prior probability shift.
A fast, non-iterative algorithm to improve computational efficiency in Section \ref{sec:algorithm}.
Section \ref{sec:simulation} presents comprehensive simulation studies, and Section \ref{sec:example} provides analyses of two empirical examples, illustrating the practical utility of the proposed method.
A brief summary and further discussion are given in Section \ref{sec:end}.

\section{IPD-AD Integrated Analysis}
\label{sec:CMLE}

\subsection{A Brief Review} 
\label{subsec:review}

We begin by reviewing the constrained maximum likelihood estimation (CMLE) framework for integrated IPD-AD analysis \citep{Qin2000Combining, ChatterjeeCMC2016Constrained}.
Let $Y$ be the outcome of interest and $\bfX$ be the $p$-dimensional covariate.
Suppose the IPD, $\{(Y_i,\bfX_i):i=1,\dots,n\}$, are independent and identically distributed copies of $(Y,\bfX)$.
We assume a parametric model, $f(y|\bfx;\bfgk{\beta})$, for the conditional density of $Y$ given $\bfX = \bfx$,  where $f$ is a known function and $\bfgk{\beta}$ is a $d$-dimensional parameter.
The marginal distribution of the covariates, $G(\bfx)$, is left unspecified and treated as a nuisance parameter.
The full log-likelihood function based on the IPD is given by $\ell(\bfgk{\beta},G)=\sum_{i=1}^n\log f(Y_i\cond\bfX_i;\bfgk{\beta})+\sum_{i=1}^n\log \Delta G(\bfX_i)$.
By a standard likelihood argument, $\ell(\bfgk{\beta},G)$ is maximized when $G$ assigns probability mass only to the observed covariate values $\{\bfX_i:i=1,\dots,n\}$.
Let $p_i$ denote the jump size of $G$ at $\bfX_i$, $i=1,\dots,n$, and define $\bfp=(p_1,\dots,p_n)$.
The log-likelihood can then be re-expressed as $\ell(\bfgk{\beta},\bfp)=\sum_{i=1}^n\log f(Y_i\cond\bfX_i;\bfgk{\beta})+\sum_{i=1}^n\log p_i$, where $p_i\ge 0$ and $\sum_{i=1}^n p_i=1$.
Denote by $(\widetilde{\bfgk{\beta}}, \widetilde{\bfp})$ the maximum likelihood estimators; in the absence of AD information, $\widetilde{\bfp}=(\widetilde{p}_1,\ldots,\widetilde{p}_n)$ is given by $\widetilde{p}_i=n^{-1}$.

As noted by several authors, including \cite{ChatterjeeCMC2016Constrained}, \cite{ZhangDSQY2020Generalized} and \cite{HuangQT2016Efficient}, AD can often be represented as an estimate  $\widetilde{\bfgk{\phi}}$ of a $q$-dimensional parameter $\bfgk{\phi}$ that satisfies the population estimating equation $\E\{\bfgk{\varphi}(Y,\bfX;\bfgk{\phi},\bfgk{\theta})\}=\bfxy{0}$, where the $r$-dimensional function $\bfgk{\varphi}$ may also depend on an additional unknown $s$-dimensional parameter $\bfgk{\theta}$.
Examples of $\bfgk{\phi}$, including subgroup-specific summary statistics, will be discussed in Section \ref{subsec:AggregateData}.
By double expectation, the parameter of interest $\bfgk{\beta}$ can be linked with the AD parameter $\bfgk{\phi}$ via the population estimating equation $\E\{\bfgk{\Psi}(\bfX;\bfgk{\beta},\bfgk{\phi},\bfgk{\theta})\}=\bfxy{0}$, where
\begin{align*}
    \bfgk{\Psi}(\bfX;\bfgk{\beta},\bfgk{\phi},\bfgk{\theta})=\E\{\bfgk{\varphi}(Y,\bfX;\bfgk{\phi},\bfgk{\theta})\cond\bfX \}=\int \bfgk{\varphi}(y,\bfX;\bfgk{\phi},\bfgk{\theta})f(y\cond\bfX;\bfgk{\beta})\diff y.
\end{align*}
When the IPD and AD follow the same distribution, the AD population moment can be approximated by the IPD sample moment, leading to the sample moment constraint $\sum_{i=1}^np_i\bfgk{\Psi}(\bfX_i;\bfgk{\beta},\bfgk{\phi},\bfgk{\theta})=\bfxy{0}$.
When the AD sample size is sufficiently large, $\widetilde{\bfgk{\phi}}$ has negligible uncertainty and may be treated as the true value of $\bfgk{\phi}$.
To integrate the AD, the constrained MLE $(\widehat{\bfgk{\beta}},\widehat{\bfp},\widehat{\bfgk{\theta}})$ is obtained by maximizing the full log-likelihood $\ell(\bfgk{\beta},\bfp)$ with respect to $(\bfgk{\beta},\bfp,\bfgk{\theta})$ subject to the constraints  
\begin{align}
    p_i\geq0,\quad\sum_{i=1}^np_i=1,\quad\sum_{i=1}^np_i\bfgk{\Psi}(\bfX_i;\bfgk{\beta},\widetilde{\bfgk{\phi}},\bfgk{\theta})=\bfxy{0}.
    \label{eq:EL_constraints_noshift}
\end{align}
Applying a standard empirical likelihood argument \citep{QinL1994Empirical, Qin2000Combining}, we have
\begin{align}
    p_i= \frac{1}{n}\times \frac{1}{1+{\bfgk{\eta}}^\T\bfgk{\Psi}(\bfX_i;{\bfgk{\beta}},\widetilde{\bfgk{\phi}},{\bfgk{\theta}})},\quad i=1,\dots,n,\label{eq:empirical_weights}
\end{align}
and the constrained log-likelihood, up to a constant,
\begin{align*}
    \sum_{i=1}^n\log f(Y_i\cond\bfX_i;\bfgk{\beta})-\sum_{i=1}^n\log\{1+\bfgk{\eta}^\T\bfgk{\Psi}(\bfX_i;\bfgk{\beta},\widetilde{\bfgk{\phi}},\bfgk{\theta})\},
\end{align*}
where ${\bfgk{\eta}}=(\eta_1,\dots,\eta_r)$ are the Lagrange multipliers satisfying
\begin{align}
    \sum_{i=1}^n\frac{\bfgk{\Psi}(\bfX_i;{\bfgk{\beta}},\widetilde{\bfgk{\phi}},{\bfgk{\theta}})}{1+\bfgk{\eta}^\T\bfgk{\Psi}(\bfX_i;{\bfgk{\beta}},\widetilde{\bfgk{\phi}},{\bfgk{\theta}})}=\bfxy{0}.\label{eq:Lagrange_equation}
\end{align}
The constrained MLE can be shown to be more efficient than its unconstrained counterpart.
Intuitively, this is because the additional constraints from the AD further restrict the parameter estimates to values that are consistent with both the IPD and the AD, thereby reducing estimation variability.

When the uncertainty in the AD cannot be ignored, we adopt the strategy proposed by \citet{ZhangDSQY2020Generalized} to treat the AD as a realization of a random vector.
Specifically, when the AD sample size $N$ is sufficiently large, the quantity $N^{1/2}(\widetilde{\bfgk{\phi}} - \bfgk{\phi})$ is approximately normal with mean zero and variance-covariance matrix $\bfgk{\Sigma}_{\bfgk{\phi}}$.
To incorporate AD, we augment the IPD log-likelihood with a quadratic term based on the normal approximation of AD:
\begin{align}
    \ell(\bfgk{\beta},\bfp,\bfgk{\phi})=\sum_{i=1}^n\log f(Y_i\cond\bfX_i;\bfgk{\beta})+\sum_{i=1}^n\log p_i-\frac{N}{2}(\widetilde{\bfgk{\phi}}-\bfgk{\phi})^\T\bfxy{V}^{-1}(\widetilde{\bfgk{\phi}}-\bfgk{\phi}),\label{eq:aug_ll}
\end{align}
where $\bfxy{V}$ is a working variance-covariance matrix. 
The constrained MLE is obtained by maximizing the augmented log-likelihood subject to the constraints
\begin{align}
    p_i\geq0,\quad\sum_{i=1}^np_i=1,\quad\sum_{i=1}^np_i\bfgk{\Psi}(\bfX_i;\bfgk{\beta},\bfgk{\phi},\bfgk{\theta})=\bfxy{0},\label{eq:EL_constraints}
\end{align}
with respect to $(\bfgk{\beta},\bfp,\bfgk{\phi},\bfgk{\theta})$.
Arguing as before, the constrained log-likelihood, up to a constant, can be reformulated as
\begin{align}
    \sum_{i=1}^n\log f(Y_i\cond\bfX_i;\bfgk{\beta})-\frac{N}{2}(\widetilde{\bfgk{\phi}}-\bfgk{\phi})^\T\bfxy{V}^{-1}(\widetilde{\bfgk{\phi}}-\bfgk{\phi})- \sum_{i=1}^n\log\{1+\bfgk{\eta}^\T\bfgk{\Psi}(\bfX_i;\bfgk{\beta},\bfgk{\phi},\bfgk{\theta})\},\label{eq:CLL_uncertain}
\end{align}
where $\bfgk{\eta}$ consists of the Lagrange multipliers satisfying
\begin{align*}
    \sum_{i=1}^n\frac{\bfgk{\Psi}(\bfX_i;{\bfgk{\beta}},\bfgk{\phi},{\bfgk{\theta}})}{1+\bfgk{\eta}^\T\bfgk{\Psi}(\bfX_i;{\bfgk{\beta},\bfgk{\phi}},{\bfgk{\theta}})}=\bfxy{0}.
\end{align*}

As shown in \citet{ZhangDSQY2020Generalized}, any positive-definite $K \times K$ choice of $\bfxy{V}$ yields an $n^{1/2}$-consistent estimator.
The optimal choice is $\bfxy{V} = \bfgk{\Sigma}_{\bfgk{\phi}}$, which leads to the most efficient estimator $\widehat{\bfgk{\beta}}$.
The asymptotic properties of $\widehat{\bfgk{\beta}}$ are summarized in the following theorem.
\begin{theorem}\label{thm:normality}
    Let $(\bfgk{\beta}_0,\bfgk{\phi}_0,\bfgk{\theta}_0,\bfxy{0})$ be the true value of $(\bfgk{\beta},\bfgk{\phi},\bfgk{\theta},\bfgk{\eta})$ and $(\widehat{\bfgk{\beta}},\widehat{\bfgk{\phi}},\widehat{\bfgk{\theta}},\widehat{\bfgk{\eta}})$ be the constrained MLE.
    Suppose that the regularity conditions A1--A6 in Supplementary Material are satisfied and assume that $N/n\to\kappa\in(0,\infty)$.
    Then, $n^{1/2}\{(\widehat{\bfgk{\beta}},\widehat{\bfgk{\phi}},\widehat{\bfgk{\theta}},\widehat{\bfgk{\eta}})-(\bfgk{\beta}_0,\bfgk{\phi}_0,\bfgk{\theta}_0,\bfxy{0})\}$ converges to a normal distribution with mean zero and variance-covariance matrix $\bfxy{J}^{-1}\bfgk{\Sigma}\bfxy{J}^{-1}$, where
    \begin{align*}
        \bfgk{\Sigma}&=\begin{pmatrix}
            -\bfxy{H}_{\bfgk{\beta}_0} & \bfxy{0} & \bfxy{0} & \bfxy{0}\\
            \bfxy{0} & \kappa\bfgk{\Sigma}_{\bfgk{\phi}}^{-1} & \bfxy{0} & \bfxy{0}\\
            \bfxy{0} & \bfxy{0} & \bfxy{0} & \bfxy{0}\\
            \bfxy{0} & \bfxy{0} & \bfxy{0} & \E\{\bfgk{\Psi}(\bfX;\bfgk{\beta}_0,\bfgk{\phi}_0,\bfgk{\theta}_0)^{\otimes2}\}
        \end{pmatrix},\\
        \bfxy{J}&=\begin{pmatrix}
            -\bfxy{H}_{\bfgk{\beta}_0} & \bfxy{0} & \bfxy{0} & \dot{\bfgk{\Psi}}_{\bfgk{\beta}}\\
            \bfxy{0} & \kappa\bfgk{\Sigma}_{\bfgk{\phi}}^{-1} & \bfxy{0} & \dot{\bfgk{\Psi}}_{\bfgk{\phi}}\\
            \bfxy{0} & \bfxy{0} & \bfxy{0} & \dot{\bfgk{\Psi}}_{\bfgk{\theta}}\\
            \dot{\bfgk{\Psi}}_{\bfgk{\beta}}^\T & \dot{\bfgk{\Psi}}_{\bfgk{\phi}}^\T & \dot{\bfgk{\Psi}}_{\bfgk{\theta}}^\T & -\E\{\bfgk{\Psi}(\bfX;\bfgk{\beta}_0,\bfgk{\phi}_0,\bfgk{\theta}_0)^{\otimes2}\}
        \end{pmatrix},
    \end{align*}
    $\bfxy{H}_{\bfgk{\beta}}=\E\{\partial^2\log f(Y\cond\bfX;\bfgk{\beta})/\partial\bfgk{\beta}^{\otimes2}\}$, and $(\dot{\bfgk{\Psi}}_{\bfgk{\beta}}^\T,\dot{\bfgk{\Psi}}_{\bfgk{\phi}}^\T,\dot{\bfgk{\Psi}}_{\bfgk{\theta}}^\T)^\T=\E\{\partial\bfgk{\Psi}(\bfX;\bfgk{\beta}_0,\bfgk{\phi}_0,\bfgk{\theta}_0)/\partial(\bfgk{\beta}^\T,\bfgk{\phi}^\T,\bfgk{\theta}^\T)^\T\}$.
\end{theorem}
\noindent As will be demonstrated in Section \ref{sec:algorithm}, the asymptotic results in Theorem \ref{thm:normality} can be used to construct an asymptotically equivalent estimator that can be efficiently obtained via a non-iterative algorithm.

\subsection{Covariate-Stratified and Outcome-Stratified Aggregate Data}
\label{subsec:AggregateData}

In this paper, we focus on the most common forms of aggregate data: subgroup summary statistics, such as means or proportions, stratified by either exposure or outcome.
These summaries are routinely reported in biomedical research, often under so-called ``Table 1," to characterize the distribution of covariates or outcomes across subpopulations and are thus readily accessible for integrated IPD-AD analysis.

One important example is the summary of outcome within subgroups defined by baseline covariates such as age, sex, or exposure status. Subgroups may be defined by a single covariate or a combination of multiple covariates, yielding a collection of (possibly overlapping) subsets $\Omega_1, \dots, \Omega_K$.
The reported subgroup means approximate the population parameters $\phi_k=\E(Y\cond\bfX\in\Omega_k)$, $k=1,\dots,K$, each of which can be reexpressed as the solution to the estimating equation $\E\{\varphi_k(Y,\bfX;\phi_k)\}=0$ with $\varphi_k(Y,\bfX;\phi_k)=(Y-\phi_k)I(\bfX\in\Omega_k)$.
By double expectation, $\phi_k$ can be further linked to the outcome model $f(y\cond\bfx;\bfgk{\beta})$ through $\E\{\Psi_k(\bfX;\bfgk{\beta},\phi_k)\} = 0$, where $\Psi_k(\bfX;\bfgk{\beta},\phi_k) = \{\int yf(y\cond\bfX;\bfgk{\beta})\diff y-\phi_k\}I(\bfX\in\Omega_k)$.
For continuous outcomes, subgroup means are often accompanied by variance estimates, which can be incorporated as additional constraints.
Intuitively, increasing the number of subgroups provides richer auxiliary information and can improve efficiency.
However, substantial overlap among subgroups induces strong correlations among summary statistics, potentially resulting in numerical instability.
Thus, careful selection is needed to balance informativeness against redundancy when selecting summaries for integration.
 
Another common form of AD summarizes covariates within subgroups defined by the outcome. A typical example is the case-control study, where covariate means or proportions are reported separately for cases ($Y=1$) and controls ($Y=0$) to highlight potential risk factors.
The outcome-stratified summary statistics approximates $\bfgk{\phi}_k=\E(\bfX\cond Y=k)$, $k=0,1$, which correspond to the solutions of the population estimating equation  $\E\{\bfgk{\varphi}_k(Y,\bfX;\bfgk{\phi}_k)\} = \bfgk{0}$, with $\bfgk{\varphi}_k(Y,\bfX;\bfgk{\phi}_k) = (\bfX -\bfgk{\phi}_k)I(Y=k)$.
By double expectation, $\bfgk{\phi}_k$ can be linked to the outcome model $\pr(Y = 1\cond\bfX = \bfx) = f(\bfx; \bfgk{\beta})$ through the estimating equation $\E\{\bfgk{\Psi}_k(\bfX;\bfgk{\beta},\bfgk{\phi}_k)\} = \bfxy{0}$, where $\bfgk{\Psi}_k(\bfX;\bfgk{\beta},\bfgk{\phi}_k) = (\bfX-\bfgk{\phi}_k)f(\bfX;\bfgk{\beta})^{k}\{1-f(\bfX;\bfgk{\beta})\}^{1-k}$.
This framework extends naturally to continuous outcomes by partitioning the outcome range into intervals,  $-\infty = y^*_0 < y^*_1 < \cdots < y^*_{K-1} < y_K = \infty$.
For each interval $(y^*_{k-1}, y^*_k]$, the conditional covariate mean $\bfgk{\phi}_k=\E(\bfxy{X}\cond y^*_{k-1}<Y\leq y^*_{k})$, $k=1,\dots,K$,  solves the estimating equation $\E\{\bfgk{\varphi}_k(Y,\bfX;\bfgk{\phi}_k)\}=\bfgk{0}$, with $\bfgk{\varphi}_k(Y,\bfX;\bfgk{\phi}_k)=(\bfX-\bfgk{\phi}_k)I(y^*_{k-1}<Y\leq y^*_k)$.
Under a parametric model $f(y\cond\bfx;\bfgk{\beta})$ for $Y$ given $\bfX$, the parameter $\bfgk{\phi}_k$  solves the population  estimating equation $\E\{\bfgk{\Psi}_k(\bfX;\bfgk{\beta},\bfgk{\phi}_k)\}=\bfxy{0}$, where  $\bfgk{\Psi}_k(\bfX;\bfgk{\beta},\bfgk{\phi}_k)=(\bfX-\bfgk{\phi}_k)\int_{y^*_{k-1}}^{y^*_k} f(y\cond\bfX;\bfgk{\beta})\diff y$.
Outcome-stratified summaries are often more transparent and less redundant than covariate-stratified AD because the one-dimensional nature of the outcome allows for systematic partitioning.
Given these conceptual advantages for integrated IPD-AD analysis, the infrequent reporting of outcome-stratified summaries for continuous outcomes presents a missed opportunity for more effective practice.

\section{IPD-AD Analysis in the Presence of Dataset Shift}
\label{sec:DatasetShift}

Dataset shift can occur when IPD and AD are drawn from different populations, resulting in differences in the joint distribution of the outcome and covariates.
For ease of discussion, we focus on the case where the AD are obtained from a single study, though the proposed framework can be extended to accommodate multiple sources.
Let $f(y, \bfx)$ represent the joint density of the random variables $(Y, \bfX)$ in the IPD study, and let $f^*(y, \bfx)$ represent the joint density of $(Y^*, \bfX^*)$ in the AD study.
Dataset shift occurs when $f(y, \bfx) \neq f^*(y, \bfx)$. 

To characterize dataset shift, we introduce a general model that encompasses two major types of shifts: covariate shift and prior probability shift.
Under covariate shift, the marginal distribution of $\bfX$ differs across studies but the conditional distribution of $Y$ given $\bfX$ remains the same.
Let $f(y\cond\bfx)$ denote the conditional density of $Y$ given $\bfX$ and let $G(\bfx)$ denote the marginal distribution function of $\bfX$; moreover, let $f^*(y\cond\bfx)$ and $G^*(\bfx)$ be defined analogously for $(Y^*, \bfX^*)$.
Write $\diff G^*(\bfx) = w_{\bfX}(\bfx)\diff G(\bfx)/\int w_{\bfX}(\bfu)\diff G(\bfu)$.
Here $\diff G^*(\bfx) /\diff G(\bfx) \propto w_{\bfX}(\bfx)$ can be viewed as a shift function characterizing the heterogeneity in the covariate distribution.
It follows from $f(y\cond\bfx)=f^*(y\cond\bfx)$ and $\int f(y \cond\bfx) \diff y =1$ that
\begin{align*}
    f^*(y,\bfx)
    = \frac{ f(y\cond\bfx) w_{\bfX}(\bfx) \diff G(\bfx)/\diff\bfx}{ \iint f(v\cond\bfu) w_{\bfX}(\bfu) \diff G(\bfu) \diff v} 
    = \frac{ w_{\bfX}(\bfx) f(y, \bfx) }{ \iint w_{\bfX}(\bfu)   f(v, \bfu) \diff v \diff \bfu}.
\end{align*}
Next, we consider prior probability shift, also referred to as label shift in the case of categorical outcomes.
In this case, the marginal distribution of $Y$ varies across studies, while the conditional distribution of $\bfX$ given $Y$ remains unchanged.
With a slight abuse of notation, let $f(\bfx\cond y)$ denote the conditional density of $\bfX$ given $Y$ and $H(y)$ denote the marginal distribution function of $Y$ in the IPD study.
Define $f^*(\bfx\cond y)$ and $H^*(y)$ analogously for $(Y^*, \bfX^*)$ in the AD study.
Write $\diff H^*(y) = w_{Y}(y)\diff H(y)/\int w_{Y}(v)\diff H(v)$; here $w_{Y}(y)$ is the shift function that characterizes heterogeneity in the distribution of $Y$ across studies.
Following $f(\bfx\cond y)=f^*(\bfx\cond y)$ and $\int f(\bfx\cond y) \diff\bfx=1$, the joint density function under prior probability shift is given by
\begin{align*}
    f^*(y,\bfx)
    = \frac{f(\bfx \cond y) w_{Y}(y) \diff H(y)/\diff y}{ \iint f(\bfu \cond v) w_{Y}(v) \diff H(v) \diff \bfu } 
    = \frac{w_{Y}(y) f(y,\bfx) }{\iint w_{Y}(v) f(v,\bfu) \diff \bfu \diff v}.
\end{align*}
Note that in this case, we have
\begin{align}
    f^*(y\cond\bfx)
    = \frac{w_{Y}(y) f(y\cond\bfx) \diff G(\bfx)}{\int w_{Y}(v) f(v\cond\bfx) \diff G(\bfx) \diff v}
    = \frac{w_{Y}(y) f(y\cond\bfx) }{\int w_{Y}(v) f(v\cond\bfx)   \diff v}  \neq f(y\cond \bfx),\label{eq:PPS_f_shifted}
\end{align}
indicating that the conditional distribution of $Y$ given $\bfX$ varies across studies. 
Finally, to accommodate both covariate shift and prior probability shift, we employ a biased-sampling formulation:
\begin{align}
    f^*(y,\bfx)
    = \frac{ w_{\bfX}(\bfx)w_{Y}(y) f(y,\bfx) }{\iint w_{\bfX}(\bfu)w_{Y}(v) f(v,\bfu) \diff \bfu \diff v}.\label{eq:generalDS}
\end{align}
In this framework, $(Y^*, \bfX^*)$ from the AD study can be viewed as a biased sample of  $(Y, \bfX)$ from the IPD study population, with sampling weights $w_{\bfX}(\bfx)w_{Y}(y)$.
When $w_{Y}(y)$ is constant, the sampling weight depends only on the covariates, corresponding to covariate shift; when $w_{\bfX}(\bfx)$ is constant, the sampling weight depends only on the outcome, corresponding to prior probability shift.

Under the general dataset shift model (\ref{eq:generalDS}), we have
\begin{align}
    \E\{\bfgk{\Psi}(\bfX^*;\bfgk{\beta},\bfgk{\phi},\bfgk{\theta})\}
    &=\E[\E\{\bfgk{\varphi}(Y^*,\bfX^*;\bfgk{\phi},\bfgk{\theta})\cond\bfX^*\}]\notag\\
    &=\frac{\int\{w_{\bfX}(\bfx)\int \bfgk{\varphi}(y,\bfx;\bfgk{\phi},\bfgk{\theta})w_{Y}(y) f(y\cond\bfx;\bfgk{\beta}) \diff y\}\diff G(\bfx)}{\iint w_{\bfX}(\bfx)w_{Y}(y) f(y\cond\bfx;\bfgk{\beta}) \diff y\diff G(\bfx)}.\label{eq:Psi_corrected}
\end{align}
Define $\bfgk{\psi}(\bfX;\bfgk{\beta},\bfgk{\phi},\bfgk{\theta})=w_{\bfX}(\bfX)\int \bfgk{\varphi}(y,\bfX;\bfgk{\phi},\bfgk{\theta})w_{Y}(y) f(y\cond\bfX;\bfgk{\beta}) \diff y$.
To obtain CMLE under dataset shift, one can maximize \eqref{eq:CLL_uncertain} by substituting 
$\bfgk{\Psi}(\bfX;\bfgk{\beta},\bfgk{\phi},\bfgk{\theta})$ with $\bfgk{\psi}(\bfX;\bfgk{\beta},\bfgk{\phi},\bfgk{\theta})$ in the constraints.
Under covariate shift, $\bfgk{\psi}(\bfX;\bfgk{\beta},\bfgk{\phi},\bfgk{\theta})$ reduces to
\begin{align}
    w_{\bfX}(\bfX)\int \bfgk{\varphi}(y,\bfX;\bfgk{\phi},\bfgk{\theta}) f(y\cond\bfX;\bfgk{\beta}) \diff y
    =w_{\bfX}(\bfX)\bfgk{\Psi}(\bfX;\bfgk{\beta},\bfgk{\phi},\bfgk{\theta}).\label{eq:correctedPsi_CS}
\end{align}
Since the marginal densities are typically assumed to be non-informative about $\bfgk{\beta}$, $w_{\bfX}$ is a constant function in $\bfgk{\beta}$.
When the AD only comprise overall covariate summaries, $\bfgk{\psi}(\bfX;\bfgk{\beta},\bfgk{\phi},\bfgk{\theta})$ does not depend on $\bfgk{\beta}$.
In this case, the maximizer of (\ref{eq:CLL_uncertain}) with respect to $\bfgk{\beta}$ is equivalent to $\widetilde{\bfgk{\beta}}=\argmax_{\bfgk{\beta}}\sum_{i=1}^n\log f(Y_i\cond\bfX_i;\bfgk{\beta})$.
Therefore, combining such type of AD yields no efficiency gain over the unconstrained MLE.
On the other hand, when the AD involve outcome-related summaries (e.g. overall outcome summaries, covariate-stratified outcome summaries, and outcome-stratified covariate summaries),  $\bfgk{\psi}(\bfX;\bfgk{\beta},\bfgk{\phi},\bfgk{\theta})$ varies with $\bfgk{\beta}$ and yield efficiency gain over $\widetilde{\bfgk{\beta}}$.
Next, under prior probability shift, $\bfgk{\psi}(\bfX;\bfgk{\beta},\bfgk{\phi},\bfgk{\theta})$ reduces to
\begin{align}
    \int \bfgk{\varphi}(y,\bfX;\bfgk{\phi},\bfgk{\theta})w_{Y}(y) f(y\cond\bfX;\bfgk{\beta}) \diff y.\label{eq:correctedPsi_PPS}
\end{align}
Consequently, most types of AD give rise to $\bfgk{\psi}(\bfX;\bfgk{\beta},\bfgk{\phi},\bfgk{\theta})$ that depends on  $\bfgk{\beta}$ and the resulting efficiency gain increases with the degree of shift between $H^*(y)$ and $H(y)$ which is characterized by $w_Y(y)$.

In practice, implementing CMLE under dataset shift poses two main challenges. 
First, the shift functions $w_{\bfX}(\bfx)$ and $w_Y(y)$ are generally unknown and must be estimated. 
Following \cite{Qin1998Inferences}, we adopt semiparametric density ratio models and jointly estimate the model parameters with $(\bfgk{\beta},\bfgk{\phi},\bfgk{\theta})$ by maximizing the constrained log-likelihood. 
For example, under covariate shift, we specify $w_{\bfX}(\bfx;\bfgk{\theta}^*)=\exp\{\bfxy{h}(\bfx)^\T\bfgk{\theta}^*\}$, where $\bfxy{h}(\cdot)$ is a pre-specified $s$-dimensional function and $\bfgk{\theta}^*$ is a $s$-dimensional parameter vector; e.g., setting $\bfxy{h}(\bfx)=\bfx$ yields exponential tilted density ratio model \citep{QinZ1997Goodness}.
Similarly, under prior probability shift, one may consider $w_{Y}(y;\bfgk{\theta}^*)=\exp\{\bfxy{h}(y)^\T\bfgk{\theta}^*\}$.
If the log density ratio between $H^*$ and $H$ is smooth, it can be approximated by a polynomial expansion using Taylor’s theorem, motivating the specification $\bfxy{h}(y)=(y,y^2,\dots,y^s)^\T$. 
While the density ratio models introduce an extra parameter $\bfgk{\theta}^*$, it only appears in the constraint (\ref{eq:EL_constraints}) and can be regarded as a component of the nusiance $\bfgk{\theta}$ in CMLE.
Therefore, the algorithms for maximizing the constrained log-likelihood (\ref{eq:CLL_uncertain}) can be directly adopted to estimate the parameters $\bfgk{\beta}$, $\bfgk{\phi}$, and $\bfgk{\theta}$, including $\bfgk{\theta}^*$.
As noted by \citet{ShengSHK2022Synthesizing}, the selection of covariates to be included in the covariate shift function $w_{\bfX}(\bfx; \bfgk{\theta}^*)$ can be guided by comparing summary statistics across studies, in conjunction with knowledge of study design and subject-matter expertise.
A similar approach can be applied to inform the model specification for $w_Y(y;\bfgk{\theta}^*)$ under prior probability shift.
These marginal summaries provide important constraints for identifying the shift model parameters $\bfgk{\theta}$.

The second challenge is the computation of $\bfgk{\psi}(\bfX;\bfgk{\beta},\bfgk{\phi},\bfgk{\theta})$, which involves an additional integration term,
\begin{align}
    \int \bfgk{\varphi}(y,\bfx;\bfgk{\phi},\bfgk{\theta})w_Y(y) f(y\cond\bfx;\bfgk{\beta}) \diff y.\label{eq:DS_additional_integration}
\end{align}
Under covariate shift, it follows from (\ref{eq:correctedPsi_CS}) that the integration reduces to $\bfgk{\Psi}(\bfX;\bfgk{\beta},\bfgk{\phi},\bfgk{\theta})$. 
Consequently, the CMLE algorithm described in Section \ref{subsec:review} can accommodate covariate shift by directly incorporating the weights $w_{\bfX}(\bfX_i)$ into the constraint $\sum_{i=1}^n p_i \bfgk{\Psi}(\bfX_i;\bfgk{\beta},\bfgk{\phi},\bfgk{\theta})=\bfxy{0}$ without requiring evaluation of a new integral.
On the other hand, under prior probability shift and using (\ref{eq:PPS_f_shifted}), the additional integration term (\ref{eq:DS_additional_integration}) can be re-expressed as
\begin{align}
    \int \bfgk{\varphi}(y,\bfx;\bfgk{\phi},\bfgk{\theta})f^*(y\cond\bfx;\bfgk{\beta})\diff y&\times \int w_{Y}(y) f(y\cond\bfx;\bfgk{\beta}) \diff y\notag\\
    =\E\{\bfgk{\varphi}(Y^*,\bfX^*;\bfgk{\phi},\bfgk{\theta})\cond\bfX^*=\bfx\} 
    &\times \int w_{Y}(y) f(y\cond\bfx;\bfgk{\beta}) \diff y,\label{eq:correctedPsi_PPS_re}
\end{align}
where $f^*(y\cond\bfx;\bfgk{\beta})=w_{Y}(y) f(y\cond\bfx;\bfgk{\beta})/\int w_{Y}(y) f(y\cond\bfx;\bfgk{\beta}) \diff y$ is the shifted conditional density.
While numerical or Monte Carlo approximations of these integrals are possible, they are computationally intensive.
However, when $f(y\cond\bfx;\bfgk{\beta})$ belongs to an exponential family and $w_Y(y)$ follows the density ratio model $\exp\{\bfxy{h}(y)^\T\bfgk{\theta}^*\}$, the shifted conditional density also belongs to the exponential family.
Specifically,  when $f(y\cond\bfx;\bfgk{\beta})$ follows the generalized linear models and $\bfxy{h}(y)=y$, both $f^*(y\cond\bfX;\bfgk{\beta})$ and the scaling factor $\int w_{Y}(y) f(y\cond\bfX;\bfgk{\beta}) \diff y$ admit simple closed forms.
Some illustrative examples are given in Table \ref{tab:Psi_example_PPS}.
In these cases, the shifted conditional distributions belong to the same class of distributions as $f(y\cond\bfx;\bfgk{\beta})$ with natural parameters shifted by a simple function of $\theta^*$.
Consequently, for most types of AD, such as the overall summaries of outcome and covariate, outcome-stratified covariate summaries, and covariate-stratified outcome summaries, the shifted expectation $\E\{\bfgk{\varphi}(Y^*,\bfX^*;\bfgk{\phi},\bfgk{\theta})\cond\bfX^*\}$ also has a simple closed form.
Therefore, computationally intensive numerical approximations are not required for these cases.

\begin{table}
    \caption{\label{tab:Psi_example_PPS}
    Examples of outcome models $f_{Y|\bfX}(y\cond\bfx)$ and their exponentially tilted versions $f^*(y\cond\bfx) \propto f(y\cond\bfx;\bfgk{\beta})\exp(\theta^* y)$, along with the corresponding normalizing constants $\int f(y\cond\bfx;\bfgk{\beta})\exp(\theta^* y)\diff y$ under different distributional assumptions; $\tilde{\bfx}=(1,\bfx^\T)^\T$.}
    \centering
    \begin{tabular}{llc}
         \toprule
         $f_{Y|\bfX}(y\cond\bfx)$   &$f^*(y\cond\bfx)$   &$\int f(y\cond\bfx;\bfgk{\beta})\exp(\theta^* y)\diff y$\\
         \midrule
         $\mathrm{Bernoulli}\{p(\bfx;\bfgk{\beta})\}$    &$\mathrm{Bernoulli}\{\mathrm{expit}(\tilde{\bfx}^\T\bfgk{\beta}+\theta^*)\}$ &\multirow{2}{*}{$\frac{1+\exp(\tilde{\bfx}^\T\bfgk{\beta}+\theta^*)}{1+\exp(\tilde{\bfx}^\T\bfgk{\beta})}$}\\
         \quad$p(\bfx;\bfgk{\beta})=\mathrm{logit}(\tilde{\bfx}^\T\bfgk{\beta})$\\[5pt]
         $\mathrm{Poisson}\{\lambda(\bfx;\bfgk{\beta})\}$    &$\mathrm{Possion}\{\lambda(\bfx;\bfgk{\beta})\exp(-\theta^*)\}$   &\multirow{2}{*}{$\exp[\lambda(\bfx;\bfgk{\beta})\{\exp(-\theta^*)-1\}]$}\\
         \quad$\lambda(\bfx;\bfgk{\beta})=\exp(\tilde{\bfx}^\T\bfgk{\beta})$\\[5pt]
         $\mathrm{Normal}\{\mu(\bfx;\bfgk{\beta}),\sigma^2\}$   &$\mathrm{Normal}\{\mu(\bfx;\bfgk{\beta})+\sigma^2\theta^*,\sigma^2\}$   &\multirow{2}{*}{$\exp\{{(2\tilde{\bfx}^\T\bfgk{\beta}_{(1)}+\sigma^2\theta^*)\theta^*}/{2}\}$}\\
         \quad$\mu(\bfx;\bfgk{\beta})=\tilde{\bfx}^\T\bfgk{\beta}_{(1)}$\\[5pt]
         $\mathrm{Gamma}\{\nu,\lambda(\bfx;\bfgk{\beta})\}$ &$\mathrm{Gamma}\{\nu,\lambda(\bfx;\bfgk{\beta})-\theta^*\}$  &\multirow{2}{*}{$\left\{\frac{\lambda(\bfx;\bfgk{\beta})}{\lambda(\bfx;\bfgk{\beta})-\theta^*}\right\}^\nu$}\\
         \quad$\lambda(\bfx;\bfgk{\beta})=\nu\exp(-\tilde{\bfx}^\T\bfgk{\beta}_{(1)})$\\
         \bottomrule
    \end{tabular}
\end{table}

\section{A Fast Non-Iterative Algorithm}
\label{sec:algorithm}

In practice, the CMLE can be obtained by solving the following nested optimization problem:
\begin{align*}
    \max_{\bfgk{\beta},\bfgk{\phi},\bfgk{\theta}} \left[ \sum_{i=1}^n \log f(Y_i\cond\bfX_i;\bfgk{\beta}) 
    - \frac{N}{2}(\widetilde{\bfgk{\phi}} - \bfgk{\phi})^\T \bfxy{V}^{-1} (\widetilde{\bfgk{\phi}} - \bfgk{\phi})
    - \min_{\bfgk{\eta}} \sum_{i=1}^n \log\{1 + \bfgk{\eta}^\T \bfgk{\psi}(\bfX_i;\bfgk{\beta},\bfgk{\phi},\bfgk{\theta})\} \right].
\end{align*}
This optimization poses two major computational challenges.
First, for a given $(\bfgk{\beta},\bfgk{\phi},\bfgk{\theta})$, the inner minimization yields the Lagrange multiplier $\widehat{\bfgk{\eta}}$, which is computed via Newton–Raphson iterations as the limit of $\{\widehat{\bfgk{\eta}}^{(m)}: m\in \mathbb{N}\}$:
\begin{align*}
    \widehat{\bfgk{\eta}}^{(m+1)} = \widehat{\bfgk{\eta}}^{(m)} 
    + \left[\sum_{i=1}^n \frac{\bfgk{\psi}(\bfX_i; \bfgk{\beta}, \bfgk{\phi}, \bfgk{\theta})^{\otimes 2}}{\{1 + \widehat{\bfgk{\eta}}^{(m)\T} \bfgk{\psi}(\bfX_i; \bfgk{\beta}, \bfgk{\phi}, \bfgk{\theta})\}^2} \right]^{-1}
    \sum_{i=1}^n \frac{\bfgk{\psi}(\bfX_i; \bfgk{\beta}, \bfgk{\phi}, \bfgk{\theta})}{1 + \widehat{\bfgk{\eta}}^{(m)\T} \bfgk{\psi}(\bfX_i; \bfgk{\beta}, \bfgk{\phi}, \bfgk{\theta})}.
\end{align*}
These iterations are guaranteed to converge when $\bfgk{0}$ lies within the convex hull of $\{\bfgk{\psi}(\bfX_i; \bfgk{\beta}, \bfgk{\phi}, \bfgk{\theta}): i = 1, \dots, n\}$.
However, the dimensionality of the gradient and Hessian matrices can be substantial, particularly when multiple AD sources are integrated.
Second, the outer maximization can also be approached via Newton–Raphson algorithms, but its convergence to the global maximum is not guaranteed \citep{HanW2013Estimation}.
To address these computational difficulties, we adopt the framework proposed by \citet{GaoC2023Noniterative} and develop a non-iterative algorithm that avoids explicit computation of the Lagrange multipliers and circumvents the need for Newton–Raphson iterations.

The algorithm is derived by comparing the influence functions of the unconstrained MLE $\widetilde{\bfgk{\beta}}$ and that of the CMLE $\widehat{\bfgk{\beta}}$.
Let $\mathcal{L}(\bfgk{\beta},\bfgk{\phi},\bfgk{\theta},\bfgk{\eta})$ be the constrained log-likelihood in (\ref{eq:CLL_uncertain}) and $\bfxy{S}(\bfgk{\beta},\bfgk{\phi},\bfgk{\theta},\bfgk{\eta})=\partial\mathcal{L}(\bfgk{\beta},\bfgk{\phi},\bfgk{\theta},\bfgk{\eta})/\partial(\bfgk{\beta}^\T,\bfgk{\phi}^\T,\bfgk{\theta}^\T,\bfgk{\eta}^\T)^\T$.
As shown in \cite{ZhangDSQY2020Generalized}, the CMLE has an asymptotic linear expansion 
\begin{align*}
    n^{1/2}(\widehat{\bfgk{\beta}}-\bfgk{\beta}_0)=(\bfxy{I}_{d},\bfxy{0},\bfxy{0},\bfxy{0})\bfxy{J}^{-1}\cdot n^{-1/2}\bfxy{S}(\bfgk{\beta}_0,\bfgk{\phi}_0,\bfgk{\theta}_0,\bfxy{0})+o_p(1),
\end{align*}
where $\bfxy{I}_{d}$ is the $d\times d$ identity matrix.
On the other hand, the MLE based only on the IPD has the asymptotic linear expansion
\begin{align*}
    n^{1/2}(\widetilde{\bfgk{\beta}}-\bfgk{\beta}_0)&=(-\bfxy{H}_{\bfgk{\beta}_0}^{-1},\bfxy{0},\bfxy{0},\bfxy{0})\cdot n^{-1/2}\bfxy{S}(\bfgk{\beta}_0,\bfgk{\phi}_0,\bfgk{\theta}_0,\bfxy{0})+o_p(1).
\end{align*}
Comparing the two asymptotic linear expansions, the asymptotically equivalent version of $\widehat{\bfgk{\beta}}$ can be derived as
\begin{align*}
    \widetilde{\bfgk{\beta}}+(\bfxy{0},\bfxy{0},\bfxy{0},\bfxy{H}_{\bfgk{\beta}_0}^{-1}\dot{\bfgk{\psi}}_{\bfgk{\beta}})\bfxy{J}^{-1}\cdot n^{-1}\bfxy{S}(\bfgk{\beta}_0,\bfgk{\phi}_0,\bfgk{\theta}_0,\bfxy{0}).
\end{align*}
Finally, by substituting the true parameters with an initial estimator $(\widetilde{\bfgk{\beta}},\widetilde{\bfgk{\phi}},\widetilde{\bfgk{\theta}})$ and noting that $\bfxy{S}(\widetilde{\bfgk{\beta}},\widetilde{\bfgk{\phi}},\widetilde{\bfgk{\theta}},\bfxy{0})=(\bfxy{0},\bfxy{0},\bfxy{0},-n^{-1}\sum_{i=1}^n\bfgk{\psi}(\bfX_i;\widetilde{\bfgk{\beta}},\widetilde{\bfgk{\phi}},\widetilde{\bfgk{\theta}})^\T)^\T$, the proposed non-iterative estimator is given by
\begin{align}
    \widehat{\bfgk{\beta}}_{\mathrm{fast}}=\widetilde{\bfgk{\beta}}+(\bfxy{0},\bfxy{0},\bfxy{0},\widetilde{\bfxy{H}}_{\bfgk{\beta}_0}^{-1}\widetilde{\dot{\bfgk{\psi}}}_{\bfgk{\beta}})\widetilde{\bfxy{J}}^{-1}\begin{pmatrix}
        \bfxy{0}\\
        \bfxy{0}\\
        \bfxy{0}\\
        -n^{-1}\sum_{i=1}^n\bfgk{\psi}(\bfX_i;\widetilde{\bfgk{\beta}},\widetilde{\bfgk{\phi}},\widetilde{\bfgk{\theta}})
    \end{pmatrix},\label{eq:betahat_fast}
\end{align}
where $\widetilde{\bfxy{H}}_{\bfgk{\beta}_0}=n^{-1}\sum_{i=1}^n\partial^2\log f(Y_i\cond\bfX_i;\widetilde{\bfgk{\beta}})/\partial\bfgk{\beta}^{\otimes2}$ and $\widetilde{\dot{\bfgk{\psi}}}_{\bfgk{\beta}}=n^{-1}\sum_{i=1}^n\partial\bfgk{\psi}(\bfX_i;\widetilde{\bfgk{\beta}},\widetilde{\bfgk{\phi}},\widetilde{\bfgk{\theta}})/\partial\bfgk{\beta}$.
Compared to the approach of \citet{GaoC2023Noniterative}, our method simultaneously accounts for both AD uncertainty and potential dataset shift between the IPD and AD sources.
Although this generalization requires an augmented matrix $\bfxy{J}$ with additional blocks involving $\kappa\bfgk{\Sigma}_{\bfgk{\phi}}^{-1}$ and $\dot{\bfgk{\psi}}_{\bfgk{\phi}}$, the enlarged matrix needs to be computed only once to obtain the updated estimator $\widehat{\bfgk{\beta}}_{\mathrm{fast}}$.
Consequently, we do not require much additional computation time when using this algorithm.

For $\bfgk{\beta}_0$ and $\bfgk{\phi}_0$, natural initial estimators are the unconstrained MLE $\widetilde{\bfgk{\beta}}$ and the observed AD $\widetilde{\bfgk{\phi}}$, respectively.
For the nuisance parameter $\bfgk{\theta}_0$, an initial estimator $\widetilde{\bfgk{\theta}}$ can be obtained using the generalized method of moments (GMM), by minimizing $\left\Vert \sum_{i=1}^n \bfgk{\psi}(\bfX_i; \widetilde{\bfgk{\beta}}, \widetilde{\bfgk{\phi}}, \bfgk{\theta}) \right\Vert^2$ over $\bfgk{\theta}$.
Importantly, the asymptotic distribution of $\widehat{\bfgk{\beta}}_{\mathrm{fast}}$ does not depend on the specific choice of initial estimators, as long as they are $n^{1/2}$-consistent.
Theoretically, the GMM procedure requires only that $\dim(\bfgk{\psi}) \geq \dim(\bfgk{\theta})$ to ensure the $n^{1/2}$-consistency of the resulting estimator.
However, when $\dim(\bfgk{\psi}) = \dim(\bfgk{\theta})$, the initial estimator $\widetilde{\bfgk{\theta}}$ satisfies $\sum_{i=1}^n \bfgk{\psi}(\bfX_i; \widetilde{\bfgk{\beta}}, \widetilde{\bfgk{\phi}}, \widetilde{\bfgk{\theta}}) = \bfxy{0}$, resulting in no improvement of $\widehat{\bfgk{\beta}}_{\mathrm{fast}}$ over $\widetilde{\bfgk{\beta}}$.
Consequently, our proposed algorithm requires the strict inequality $\dim(\bfgk{\psi}) > \dim(\bfgk{\theta})$.
In settings with dataset shift, if the shift function is modeled flexibly such that $\dim(\bfgk{\theta}_0)$ is large, the corresponding parameter dimension $\dim(\bfgk{\theta})$ also increases.
In such cases, a greater number of aggregate summary statistics are needed to ensure that $\dim(\bfgk{\psi}) > \dim(\bfgk{\theta})$ holds.

From the asymptotic normality of $n^{-1}\bfxy{S}(\bfgk{\beta}_0,\bfgk{\phi}_0,\bfgk{\theta}_0,\bfxy{0})$ and the consistency of $\widetilde{\bfxy{H}}_{\bfgk{\beta}_0}$, $\widetilde{\dot{\bfgk{\psi}}}_{\bfgk{\beta}}$, and $\widetilde{\bfxy{J}}$, the Slutsky's theorem ensures the same asymptotic normality of $\widehat{\bfgk{\beta}}_{\mathrm{fast}}$ as $\widehat{\bfgk{\beta}}$, which is given in Theorem \ref{thm:normality}.
With the proposed estimator $\widehat{\bfgk{\beta}}_{\mathrm{fast}}$, the asymptotic variance-covariance matrix can be estimated by $\widehat{\bfxy{J}}^{-1}\widehat{\bfgk{\Sigma}}\widehat{\bfxy{J}}^{-1}$, where $\widehat{\bfxy{J}}$ is the counterpart of $\widetilde{\bfxy{J}}$ with $\widetilde{\bfgk{\beta}}$ replaced by $\widehat{\bfgk{\beta}}_{\mathrm{fast}}$, and $\widehat{\bfgk{\Sigma}}=n^{-1}\sum_{i=1}^n\bfxy{S}(\widehat{\bfgk{\beta}}_{\mathrm{fast}},\widetilde{\bfgk{\phi}},\widetilde{\bfgk{\theta}},\bfxy{0})^{\otimes2}$.
A $100(1-\alpha)\%$ confidence set for $\bfgk{\bfgk{\beta}}_0$ is constructed as $\{\bfgk{\beta}:(\widehat{\bfgk{\beta}}_{\mathrm{fast}}-\bfgk{\beta})^\T\widehat{\bfxy{J}}\widehat{\bfgk{\Sigma}}^{-1}\widehat{\bfxy{J}}(\widehat{\bfgk{\beta}}_{\mathrm{fast}}-\bfgk{\beta})\leq\chi^2_d(1-\alpha)\}$, where $\chi^2_d(1-\alpha)$ is the $(1-\alpha)$th quantile of $\chi^2_d$ distribution.
The proposed estimation and corresponding inference can be implemented by our developed R package \texttt{GLMaggregate}.

\section{Simulation Studies}
\label{sec:simulation}

\subsection{Integrated IPD-AD Analysis in the Absence of Dataset Shift}

We begin by evaluating the performance of the CMLE for integrated IPD-AD analysis in the absence of dataset shift.
For both IPD and AD, the covariate $X_1$ is generated from a standard normal distribution, and $X_2$ is generated from a Bernoulli distribution with success probability $\pr(X_2=1)=0.6$.
The response variable $Y$ is generated from a normal distribution with mean $\beta_{00}+\beta_{01}X_1+\beta_{02}X_2$ and variance $\sigma_0^2$, where $(\beta_{00},\beta_{01},\beta_{02},\sigma)=(0.5,-0.5,0.5,1)$.
We consider different sets of AD that estimate the following quantities:
\begin{align*}
    &\bfgk{\phi}^Y:\E(Y);\\
    &\bfgk{\phi}^{\bfX|Y}_{\mathrm{median}}:\E(\bfX\cond Y\leq q_{0.5}),  \; \E(\bfX\cond Y> q_{0.5});\\
    &\bfgk{\phi}^{\bfX|Y}_{\mathrm{quartile}}:\E(\bfX\cond Y\leq q_{0.25}), \; \E(\bfX\cond q_{0.25}<Y\leq q_{0.5}),  \;  \E(\bfX\cond q_{0.5}<Y\leq q_{0.75}), \; \E(\bfX\cond Y> q_{0.75});\\
    &\bfgk{\phi}^{Y|\bfX}_{1}:\E(Y\cond-1<X_1\leq 1);\\
    &\bfgk{\phi}^{Y|\bfX}_{2}:\E(Y\cond-1<X_1\leq 1),  \; \E(Y\cond X_2=1);\\
    &\bfgk{\phi}^{Y|\bfX}_{3}:\E(Y\cond-1<X_1\leq 1),  \; \E(Y\cond X_2=1),\E(Y\cond X_2=0),
\end{align*}
where $q_p$ denotes the $p$th quantile of $Y$.
The AD sample size is set to be $N=1000$.
The following simulation results are based on $1000$ replications.

\begin{figure}
    \centering
    \includegraphics[scale=0.8]{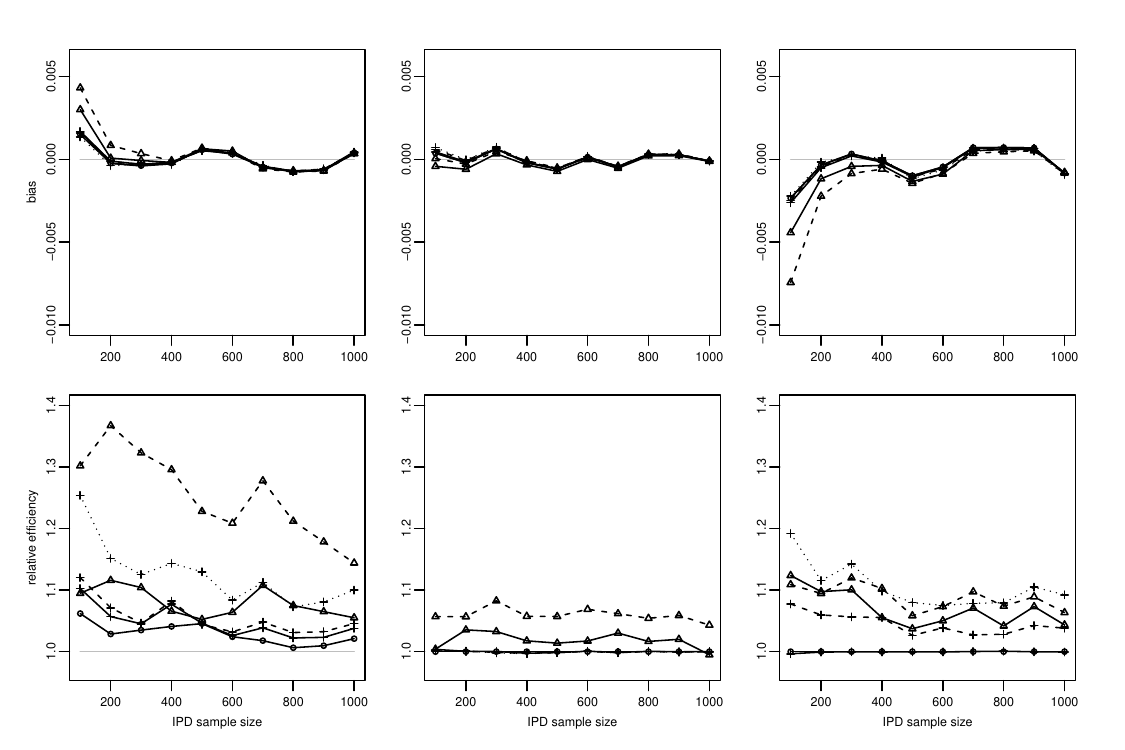}
    \caption{
    The biases (top panel) and relative efficiencies (bottom panel) of the constrained maximum likelihood estimator for $\beta_{00}$ (left), $\beta_{01}$ (center), and $\beta_{02}$ (right), with various AD: $\widetilde{\bfgk{\phi}}^Y$ (solid line with $\circ$), $\widetilde{\bfgk{\phi}}^{\bfX|Y}_{\mathrm{median}}$ (solid line with $\triangle$), $\widetilde{\bfgk{\phi}}^{\bfX|Y}_{\mathrm{quartile}}$ (dashed line with $\triangle$), $\widetilde{\bfgk{\phi}}^{Y|\bfX}_{1}$ (solid line with $+$), $\widetilde{\bfgk{\phi}}^{Y|\bfX}_{2}$ (dashed line with $+$), and $\widetilde{\bfgk{\phi}}^{Y|\bfX}_{3}$ (dotted line with $+$) under AD sample size $N=1000$.}
    \label{fig:NS_bias_RE_n}
\end{figure}

Figure~\ref{fig:NS_bias_RE_n} depicts the biases of the CMLE and its relative efficiencies compared with the unconstrained MLE.
As expected, biases are close to zero and decrease as the IPD sample size increases. 
In addition, CMLE estimators typically have smaller standard deviations than the constrained MLE, demonstrating efficiency gains by leveraging aggregate data.
Moreover, comparing the standard deviations of the CMLE estimators that incorporate covariate-stratified outcome means $\widetilde{\bfgk{\phi}}^{Y|\bfX}_{1}$, $\widetilde{\bfgk{\phi}}^{Y|\bfX}_{2}$, and $\widetilde{\bfgk{\phi}}^{Y|\bfX}_{3}$, we observe that integrating more AD leads to improved efficiency, consistent with the theoretical results.
For outcome-stratified covariate means, the CMLE estimator based on $\widetilde{\bfgk{\phi}}^{\bfX|Y}_{\mathrm{quartile}}$ outperforms that based on $\widetilde{\bfgk{\phi}}^{\bfX|Y}_{\mathrm{median}}$.
Although $\widetilde{\bfgk{\phi}}^{\bfX|Y}_{\mathrm{median}}$ is not contained in $\widetilde{\bfgk{\phi}}^{\bfX|Y}_{\mathrm{quartile}}$, the finer partitioning of the outcome range provides more detailed information about the relationship between the covariates and the outcome, and hence yields the better efficiency of combining $\widetilde{\bfgk{\phi}}^{\bfX|Y}_{\mathrm{quartile}}$ over $\widetilde{\bfgk{\phi}}^{\bfX|Y}_{\mathrm{median}}$.

\begin{figure}
    \centering
    \includegraphics[scale=0.8]{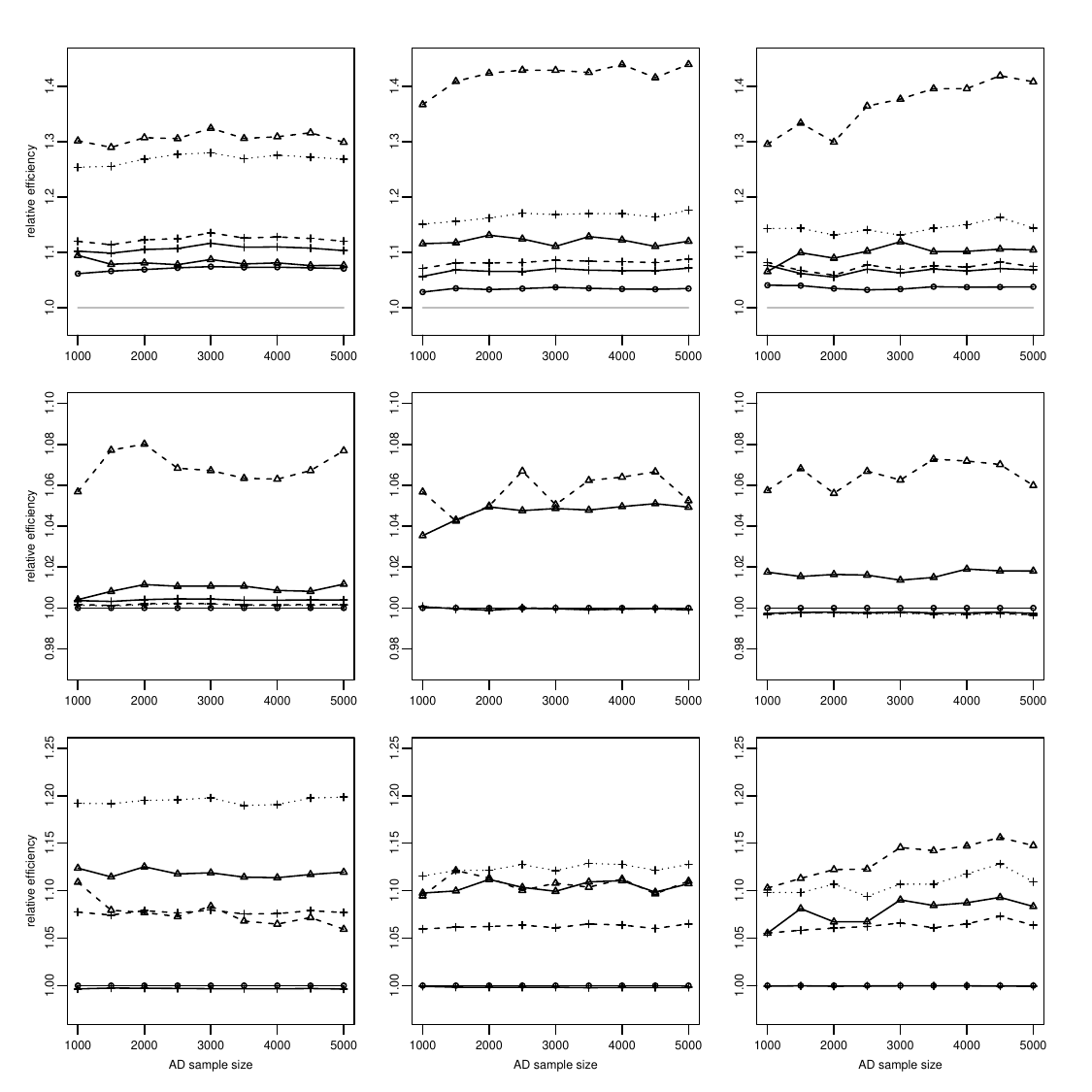}
    \caption{The relative efficiencies of the constrained maximum likelihood estimator for $\beta_{00}$ (top row), $\beta_{01}$ (center row), and $\beta_{02}$ (bottom row) under IPD sample sizes $n=100$ (left column), $n=200$ (center column), and $n=400$ (right column), with various AD: $\widetilde{\bfgk{\phi}}^Y$ (solid line with $\circ$), $\widetilde{\bfgk{\phi}}^{\bfX|Y}_{\mathrm{median}}$ (solid line with $\triangle$), $\widetilde{\bfgk{\phi}}^{\bfX|Y}_{\mathrm{quartile}}$ (dashed line with $\triangle$), $\widetilde{\bfgk{\phi}}^{Y|\bfX}_{1}$ (solid line with $+$), $\widetilde{\bfgk{\phi}}^{Y|\bfX}_{2}$ (dashed line with $+$), and $\widetilde{\bfgk{\phi}}^{Y|\bfX}_{3}$ (dotted line with $+$).}
    \label{fig:NS_RE_n_N}
\end{figure}

In general, combining outcome-stratified covariate means $\widetilde{\bfgk{\phi}}^{\bfX|Y}_{\mathrm{quartile}}$ yields greater efficiency gains than combining covariate-stratified subgroup outcome means $\widetilde{\bfgk{\phi}}^{Y|\bfX}_{3}$, except when estimating the coefficient of the discrete covariate.
The reduced efficiency is likely explained by the higher variability of $\widetilde{\bfgk{\phi}}^{\bfX|Y}_{\mathrm{quartile}}$ due to small subgroup sample sizes.
However, when the IPD sample size is sufficiently large ($n=400$) and the AD sample size increases, the bottom-right panel of Figure~\ref{fig:NS_RE_n_N} shows that the efficiency of estimators incorporating $\widetilde{\bfgk{\phi}}^{\bfX|Y}_{\mathrm{quartile}}$ ultimately outperforming other AD.

\subsection{Integrated IPD-AD Analysis Under Dataset Shift}

Next, we examine the performance of the CMLE under covariate shift or prior probability shift.
In the scenarios with covariate shift, the external covariates $\bfX^*$ are generated from the density ratio model $\diff G^*(\bfx)/\diff G(\bfx)=\exp(\theta_{01}x_1+\theta_{02}x_2)/\int\exp(\theta_{01}x_1+\theta_{02}x_2)\diff G(\bfx)$ with $(\theta_{01},\theta_{02})=(0.5,0.5)$.
In the scenarios with prior probability shift, the external outcomes $Y^*$ are generated using the density ratio model $\diff F^*(y)=\exp(\theta_0 y)\diff F(y)/\int\exp(\theta_0 y)\diff F(y)$ with $\theta_0=0.5$.
Additionally, we consider the AD corresponding to the target parameter:
\begin{align*}
    \bfgk{\phi}^{\bfX}:\E(\bf{X}^*).
\end{align*}
As discussed in Section \ref{sec:DatasetShift}, $\widetilde{\bfgk{\phi}}^X$ does not provide any information about $\bfgk{\beta}$ under covariate shift and offers little information about $\bfgk{\beta}$ when the prior probability shift is small.

\begin{figure}
    \centering
    \includegraphics[scale=0.7]{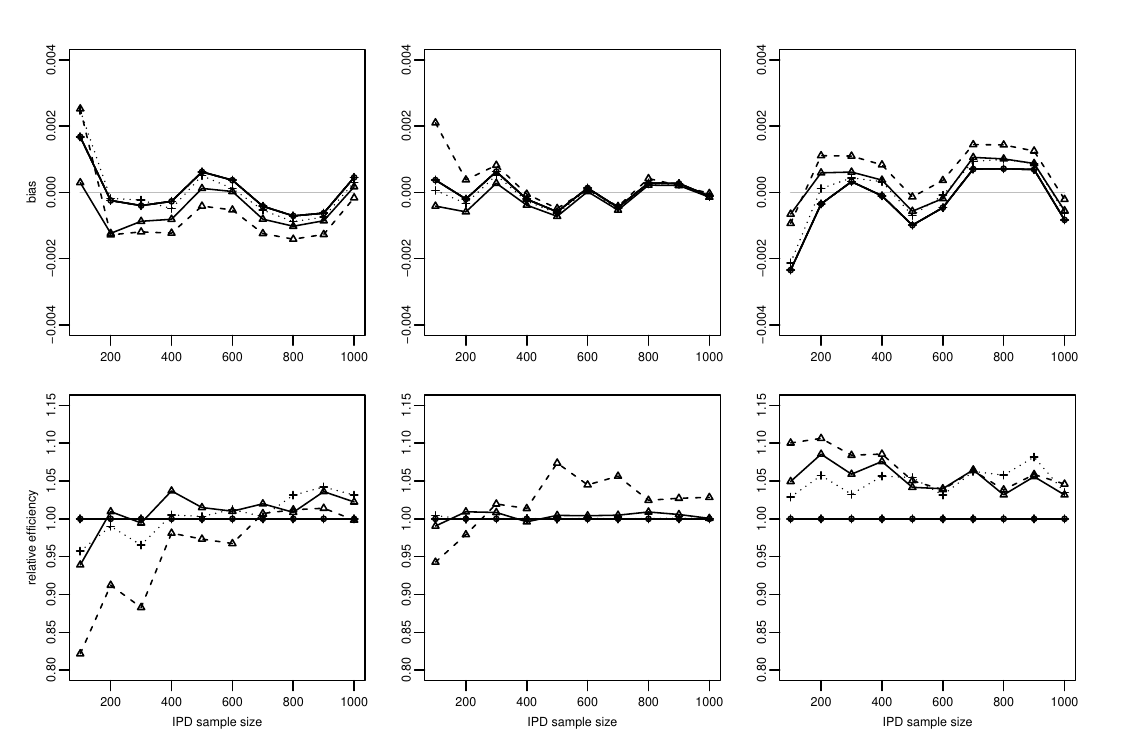}
    \caption{
    The biases (top panel) and relative efficiencies (bottom panel) of the constrained maximum likelihood estimator for $\beta_{00}$ (left), $\beta_{01}$ (center), and $\beta_{02}$ (right) under covariate shift, with various AD: $\widetilde{\bfgk{\phi}}^Y$ (solid line with $\circ$), $\widetilde{\bfgk{\phi}}^{\bfX|Y}_{\mathrm{median}}$ (solid line with $\triangle$), $\widetilde{\bfgk{\phi}}^{\bfX|Y}_{\mathrm{quartile}}$ (dashed line with $\triangle$), $\widetilde{\bfgk{\phi}}^{Y|\bfX}_{1}$ (solid line with $+$), $\widetilde{\bfgk{\phi}}^{Y|\bfX}_{2}$ (dashed line with $+$), and $\widetilde{\bfgk{\phi}}^{Y|\bfX}_{3}$ (dotted line with $+$) under AD sample size $N=1000$.}
    \label{fig:CS_bias_RE_n}
\end{figure}

\begin{figure}
    \centering
    \includegraphics[scale=0.7]{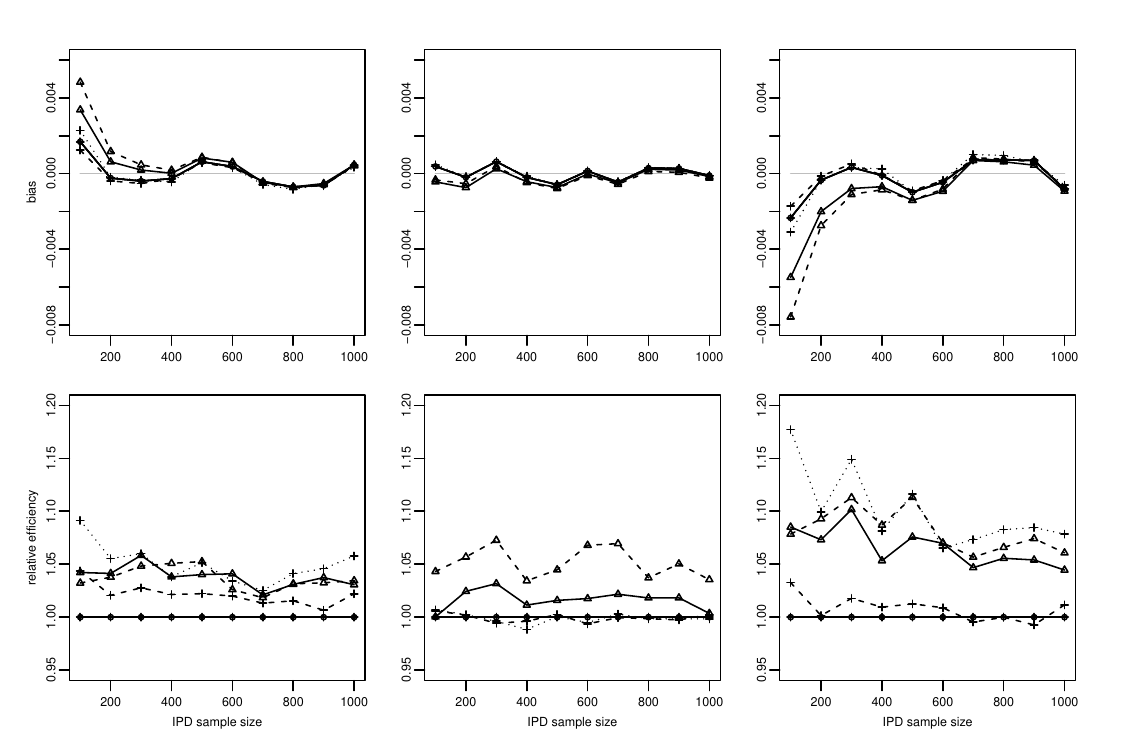}
    \caption{ 
    The biases (top panel) and relative efficiencies (bottom panel) of the constrained maximum likelihood estimator for $\beta_{00}$ (left), $\beta_{01}$ (center), and $\beta_{02}$ (right) under prior probability shift, with various AD: $\widetilde{\bfgk{\phi}}^Y$ (solid line with $\circ$), $\widetilde{\bfgk{\phi}}^{\bfX|Y}_{\mathrm{median}}$ (solid line with $\triangle$), $\widetilde{\bfgk{\phi}}^{\bfX|Y}_{\mathrm{quartile}}$ (dashed line with $\triangle$), $\widetilde{\bfgk{\phi}}^{Y|\bfX}_{1}$ (solid line with $+$), $\widetilde{\bfgk{\phi}}^{Y|\bfX}_{2}$ (dashed line with $+$), and $\widetilde{\bfgk{\phi}}^{Y|\bfX}_{3}$ (dotted line with $+$) under AD sample size $N=1000$.}
    \label{fig:PPS_bias_RE_n}
\end{figure}

The biases and relative efficiencies (relative to the unconstrained MLE) of the CMLE are presented in Figure~\ref{fig:CS_bias_RE_n} (under covariate shift) and Figure~\ref{fig:PPS_bias_RE_n} (under prior probability shift). 
Consistent with the conclusions drawn from scenarios without dataset shift, most biases are close to zero and diminish as the sample size increases.
However, we observe substantial biases when incorporating $\widetilde{\bfgk{\phi}}^{\bfX}$ under scenarios with prior probability shift. 
Although the moment condition is derived as a function of $\bfgk{\beta}$, it is very flat under the simulation settings, leading to a nearly singular $\widetilde{\bfxy{J}}_{\bfxy{V}}$ and unstable estimates. 
As a result, $\widetilde{\bfgk{\phi}}^{\bfX}$ provides limited information even in the presence of prior probability shift.

In the presence of dataset shift, incorporating the outcome-stratified subgroup covariate means and the covariate-stratified subgroup outcome means yield comparable efficiency and outperfom incorporating the marginal outcome means.
When the sample size is sufficiently large ($n\geq800$), the AD $\widetilde{\bfgk{\phi}}^{\bfX|Y}_{\mathrm{quartile}}$ and $\widetilde{\bfgk{\phi}}^{Y|\bfX}_{3}$ generally lead to more improvement in efficiency among the outcome-stratified subgroup covariate means and the covariate-stratified subgroup outcome means, respectively.
This confirms the asymptotic properties of the CMLE.
In addition, when estimating the coefficients of continuous covariates, $\widetilde{\bfgk{\phi}}^{\bfX|Y}_{\mathrm{quartile}}$ outperforms $\widetilde{\bfgk{\phi}}^{Y|\bfX}_{3}$.

\section{Empirical Examples}
\label{sec:example}

\subsection{Analysis of National Longitudinal Survey Data}
\label{subsec:NLSY}

The National Longitudinal Survey of Youth (NLSY) is a long-term study that tracks a cohort of American youth born between 1980 and 1984. 
In this subsection, we utilize this cohort data with 5095 respondents to examine the influence of demographic characteristics on income ($Y$, in thousand USD).
A fit of linear regression model to the centralized and standardized outcome:
\begin{align*}
    \tilde{Y}=\beta_0+\beta_1X_1+\cdots+\beta_5X_5+\varepsilon,
\end{align*}
where $\tilde{Y}=(Y-51.5)/33.1$ and $\varepsilon\sim N(0,\sigma^2_{\text{S}})$, identifies several significant factors: gender ($X_1=1$ for male and $X_1=0$ for female), race ($(X_2,X_3)=(1,0)$ for Hispanic, $(X_2,X_3)=(0,1)$ for Black, and $(X_2,X_3)=(0,0)$ otherwise), education ($X_4=1$ for college degree and $X_4=0$ otherwise), and residence ($X_5=1$ for urban and $X_5=0$ for non-urban). 
To improve estimation efficiency, we incorporate external AD from the 1979 wave of the NLSY and assess which types of AD contribute most to efficiency gains.
The AD sample size is 4052.
Consistent with our simulation study, the following types of AD are considered:
\begin{enumerate}
    \item $\widetilde{\bfgk{\phi}}^{\bfX}$: marginal means of the covariates;
    \item $\widetilde{\bfgk{\phi}}^{Y}$: marginal mean of the outcome (income);
    \item $\widetilde{\bfgk{\phi}}^{\bfX|Y}_{\mathrm{median}}$: subgroup means of the covariates, with subgroups defined by the median of the outcome;
    \item $\widetilde{\bfgk{\phi}}^{\bfX|Y}_{\mathrm{quartile}}$: subgroup means of the covariates, with subgroups defined by the quartiles of the outcome;
    \item $\widetilde{\bfgk{\phi}}^{Y|\bfX}$: subgroup means of the outcome, with subgroups defined by the medians of the covariates.
\end{enumerate}
The values of these summary statistics are presented in Table~\ref{tab:NLS_aggregate}.

A major challenge in integrating these AD is the presence of covariate shift. 
In the NLSY97 cohort, the proportion of male respondents is higher ($0.512$) compared to the NLSY79 cohort ($0.478$). 
Similarly, the proportion of respondents with a college degree is greater in the NLSY97 cohort ($0.652$) than in the NLSY79 cohort ($0.546$). 
We account for this covariate shift within the CMLE framework. 
The analysis results, including the estimates (Est.), their corresponding standard errors (S.E.), and relative efficiencies (R.E.) with respect to the MLE, are presented in Table~\ref{tab:NLS}. 
All coefficient estimates are comparable and exhibit relative efficiencies greater than one when incorporating different types of AD. 
In particular, combining $\widetilde{\bfgk{\phi}}^{\bfX|Y}_{\mathrm{quartile}}$ yields the smallest standard errors among the various AD types, demonstrating the effectiveness of the proposed approach.

According to the analysis results incorporating $\widetilde{\bfgk{\phi}}^{\bfX|Y}_{\mathrm{quartile}}$, male respondents earn, on average, 16.5 thousand dollars more than female respondents (95\% confidence interval [CI]: 15.1--18.0).
Respondents with a college degree earn 20.4 thousand dollars more than those without a degree (95\% CI: 18.9--21.9).
Those residing in urban areas earn 4.1 thousand dollars more than their non-urban counterparts (95\% CI: 2.3--5.8).
Hispanic respondents earn 5.4 thousand dollars less (95\% CI: 3.6--7.3), and Black respondents earn 12.2 thousand dollars less (95\% CI: 10.4--13.9) than others.

\begin{table}
    \caption{\label{tab:NLS_aggregate}
    Summary statistics from 1979 National Longitudinal Survey Data.
    $Q_1$, $Q_2$, and $Q_3$ are quartiles of outcome, and $q_{\alpha}(\cdot)$ is the $\alpha$th quantile function.}
    \centering
    \setlength{\tabcolsep}{2.5pt}
    \renewcommand{\arraystretch}{0.8}
    \footnotesize
    \begin{tabular}{lrrrrrrrrrr}
    \toprule
         & \multicolumn{7}{c}{Mean of covariates}                                                      &       & \multicolumn{2}{c}{Mean of outcome}            \\
         \cmidrule(lr){2-8}\cmidrule(lr){9-11}
Subgroup & All  & $(0, Q_2]$ & $(Q_2,\infty)$ & $(0,Q_1]$ & $(Q_1,Q_2]$ & $(Q_2,Q_3]$ & $(Q_3,\infty)$ & All   & $(0,q_{0.5}(X_j^*)]$ & $(q_{0.5}(X^*),\infty)$ \\
\midrule
         &      &            &                &           &             &             &                & 54.82 &                      &                         \\
Gender   & 0.48 & 0.38       & 0.58           & 0.36      & 0.40        & 0.54        & 0.62           &       & 63.97                & 46.45                   \\
Hispanic & 0.19 & 0.20       & 0.18           & 0.18      & 0.22        & 0.18        & 0.17           &       & 52.28                & 55.42                   \\
Black    & 0.28 & 0.33       & 0.23           & 0.35      & 0.31        & 0.26        & 0.20           &       & 46.82                & 57.97                   \\
College  & 0.55 & 0.43       & 0.67           & 0.41      & 0.45        & 0.59        & 0.74           &       & 64.72                & 42.91                   \\
Urban    & 0.79 & 0.78       & 0.80           & 0.77      & 0.80        & 0.79        & 0.81           &       & 55.62                & 51.80                  \\
\bottomrule
\end{tabular}
\end{table}

\begin{table}
    \caption{\label{tab:NLS}
    The estimated regression coefficients (Est.), corresponding standard errors (S.E.), and relative efficiencies (R.E.) in the National Longitudinal Survey data analysis}
    
    \centering
    \setlength{\tabcolsep}{2.5pt}
    \renewcommand{\arraystretch}{0.8}
    \footnotesize
\begin{tabular}{lrrrrrrrrrrrrrrr}
\toprule
          & \multicolumn{3}{c}{$\widetilde{\bfgk{\phi}}^{\bfX}$} & \multicolumn{3}{c}{$\widetilde{\bfgk{\phi}}^{Y}$} & \multicolumn{3}{c}{$\widetilde{\bfgk{\phi}}^{\bfX|Y}_{\mathrm{median}}$} & \multicolumn{3}{c}{$\widetilde{\bfgk{\phi}}^{\bfX|Y}_{\mathrm{quartile}}$} & \multicolumn{3}{c}{$\widetilde{\bfgk{\phi}}^{Y|\bfX}$} \\
          \cmidrule(lr){2-4}\cmidrule(lr){5-7}\cmidrule(lr){8-10}\cmidrule(lr){11-13}\cmidrule(lr){14-16}
          & Est.              & S.E.            & R.E.            & Est.             & S.E.           & R.E.           & Est.               & S.E.              & R.E.             & Est.               & S.E.              & R.E.             & Est.              & S.E.             & R.E.             \\
          \midrule
Intercept & -0.64             & 0.03            & 1.23            & -0.62            & 0.04           & 1.00           & -0.64              & 0.03              & 1.23             & -0.63              & 0.03              & 1.36             & -0.55             & 0.02             & 4.08             \\
Gender    & 0.52              & 0.02            & 1.27            & 0.49             & 0.03           & 1.00           & 0.52               & 0.02              & 1.27             & 0.50               & 0.02              & 1.37             & 0.47              & 0.02             & 1.15             \\
Hispanic  & -0.18             & 0.03            & 1.26            & -0.18            & 0.03           & 1.00           & -0.18              & 0.03              & 1.26             & -0.16              & 0.03              & 1.38             & -0.19             & 0.03             & 1.03             \\
Black     & -0.38             & 0.03            & 1.30            & -0.39            & 0.03           & 1.00           & -0.38              & 0.03              & 1.30             & -0.37              & 0.03              & 1.35             & -0.40             & 0.03             & 1.04             \\
College   & 0.62              & 0.02            & 1.31            & 0.60             & 0.03           & 1.00           & 0.62               & 0.02              & 1.31             & 0.62               & 0.02              & 1.46             & 0.58              & 0.02             & 1.21             \\
Urban     & 0.13              & 0.03            & 1.31            & 0.13             & 0.03           & 1.00           & 0.13               & 0.03              & 1.31             & 0.12               & 0.03              & 1.61             & 0.10              & 0.03             & 1.39            \\
\bottomrule
\end{tabular}
\end{table}

\subsection{Analysis of Japan Real Estate Price}
\label{subsec:JapanHouse}

In this subsection, we analyze real estate transaction records in Osaka, collected by Japan's Ministry of Land, Infrastructure, Transport, and Tourism. 
Our objective is to assess the impact of various factors on house prices ($Y$, in ten million yen) in Osaka's housing market. 
The considered covariates include time to the nearest station ($X_1$, in minutes), land area ($X_2$, in $\text{m}^2$), floor area ($X_3$, in $\text{m}^2$), house age ($X_4$, in years), breadth ($X_5$, in $\text{m}$), coverage ratio ($X_6$), and floor area ratio ($X_7$).
The Ministry regularly publishes summary statistics and charts related to house prices and their associated attributes.
To investigate which types of summary statistics improve the efficiency of evaluating the effects of these attributes, we use individual-level data with 1565 records from 2019 as the IPD and integrate it with summary statistics derived from the 2018 data.
The AD sample size is 2506.
Consistent with our simulation study, the considered summary statistics include both marginal means and subgroup means of the response and covariates.
Subgroups are defined based on the quantile values of the response and covariates.
The values of these statistics are presented in Table~\ref{tab:JapanHousePrice_aggregate}.

As noted by \cite{HuangQH2024Efficient}, transaction records reflect only properties that were sold and do not constitute a random sample of the entire housing stock. 
In particular, these records may be biased toward lower-priced transactions, as higher-priced homes are sold less frequently due to a smaller pool of potential buyers. 
Consequently, a prior probability shift naturally arises between datasets collected across different time periods.
We first fit a linear model to the centralized and standardized transaction prices in 2019
\begin{align*}
    \widetilde{Y} 
    =\theta_0+\theta_1X_1+\dots+\theta_7X_7+\varepsilon,
\end{align*}
where $\widetilde{Y}=(Y-25.5)/16.3$.
Here, $\varepsilon$ follows a normal distribution with mean zero and variance $\theta_8^2$ and is assumed to be independent of $\bfX$.
Subsequently, we account for this shift in the five sets of AD as described in Section \ref{subsec:NLSY} within the CMLE framework.
The analysis results, including the estimates (Est.), their corresponding standard errors (S.E.), and relative efficiencies (R.E.) with respect to the MLE, are presented in Table~\ref{tab:JapanHousePrice}. 
All regression coefficient estimates are comparable across combining different AD. 
Notably, integrating $\widetilde{\bfgk{\phi}}^{\bfX|Y}_{\mathrm{quartile}}$ generally yields the smallest standard errors among the various types of AD, with the exception of the intercept and the coefficients for house age and coverage ratio. 
This result underscores the utility of outcome-stratified covariate means as a form of AD, consistent with the findings from our simulation studies.

The fitted models suggest that distance to the nearest station, age of the house, and floor-area ratio negatively impact house price.
Notably, a one-minute increase in walking distance reduces the average price by 311.4 thousand Yen (95\% confidence interval [CI]: 257.2--365.8), while each year of age decreases the price by 482.7 thousand Yen (95\% CI: 458.4--507.0).
In contrast, land area, floor area, breadth, and coverage ratio positively affect price.
Interestingly, each additional square meter of floor area increases price by an average of 8.6 million Yen (95\% CI: 6.5--10.8), whereas the same increase in land area leads to a lower price increase of 7.1 million Yen (95\% CI: 6.2--8.0).

\begin{table}
    \caption{\label{tab:JapanHousePrice_aggregate}
    Summary statistics from 2018 Japan house price data.
    $Q_1$, $Q_2$, and $Q_3$ are quartiles of outcome, and $q_{\alpha}(\cdot)$ is the $\alpha$th quantile function.}
    \centering
    \setlength{\tabcolsep}{2.5pt}
    \renewcommand{\arraystretch}{0.8}
    \footnotesize
    \begin{tabular}{lrrrrrrrrrr}
        \toprule
                     & \multicolumn{7}{c}{Mean of covariates}                                                         &     & \multicolumn{2}{c}{Mean of outcome}                           \\
                     \cmidrule(lr){2-8}\cmidrule(lr){9-11}
        Subgroup         & All   & $(0, Q_2]$ & $(Q_2,\infty)$ & $(0,Q_1]$ & $(Q_1,Q_2]$ & $(Q_2,Q_3]$ & $(Q_3,\infty)$ & All & $(0,q_{0.5}(X_j^*)]$ & $(q_{0.5}(X_j^*),\infty)$ \\
        \midrule
                         &       &               &          &           &                  &                  &           &2.68&  &  \\
        Time             & 15.30 & 16.22         & 14.36    & 15.86     & 16.58            & 15.40            & 13.31     &     & 2.87                          & 2.50                          \\
        Land area        & 1.23  & 1.11          & 1.35     & 0.92      & 1.30             & 1.23             & 1.47      &     & 2.23                          & 3.17                          \\
        Floor area       & 1.00  & 0.92          & 1.09     & 0.81      & 1.02             & 1.03             & 1.15      &     & 2.20                          & 3.38                          \\
        Age              & 18.70 & 30.76         & 6.42     & 39.56     & 21.99            & 6.39             & 6.45      &     & 3.59                          & 1.77                          \\
        Breadth          & 5.51  & 5.30          & 5.73     & 5.11      & 5.49             & 5.62             & 5.84      &     & 2.47                          & 2.91                          \\
        Coverage ratio   & 0.60  & 0.59          & 0.61     & 0.60      & 0.59             & 0.60             & 0.61      &     & 2.63                          & 3.11                          \\
        Floor area ratio & 1.85  & 1.84          & 1.86     & 1.90      & 1.78             & 1.86             & 1.85      &     & 2.66                          & 3.07                          \\
                     \bottomrule
    \end{tabular}
\end{table}

\begin{table}
    \caption{\label{tab:JapanHousePrice}
    The estimated regression coefficients (Est.), corresponding standard errors (S.E.), and relative efficiencies (R.E.) in the Japan house price data analysis}
    \centering
    \setlength{\tabcolsep}{2.5pt}
    \renewcommand{\arraystretch}{0.8}
    \footnotesize
\begin{tabular}{lrrrrrrrrrrrrrrr}
\toprule
                 & \multicolumn{3}{c}{$\widetilde{\bfgk{\phi}}^{\bfX}$} & \multicolumn{3}{c}{$\widetilde{\bfgk{\phi}}^{Y}$} & \multicolumn{3}{c}{$\widetilde{\bfgk{\phi}}^{\bfX|Y}_{\mathrm{median}}$} & \multicolumn{3}{c}{$\widetilde{\bfgk{\phi}}^{\bfX|Y}_{\mathrm{quartile}}$} & \multicolumn{3}{c}{$\widetilde{\bfgk{\phi}}^{Y|\bfX}$} \\
                 \cmidrule(lr){2-4}\cmidrule(lr){5-7}\cmidrule(lr){8-10}\cmidrule(lr){11-13}\cmidrule(lr){14-16}
                 & Est.                 & S.E.                & R.E.               & Est.                & S.E.               & R.E.              & Est.                  & S.E.                 & R.E.                 & Est.                  & S.E.                 & R.E.                 & Est.                  & S.E.                & R.E.                \\
                 \midrule
Intercept        & -0.58                & 0.16                & 1.01               & -0.58               & 0.16               & 1.00              & -0.65                 & 0.15                 & 1.15                 & -0.67                 & 0.14                 & 1.21                 & -0.69                 & 0.14                & 1.24                \\
Time             & -0.02                & 0.00                & 1.01               & -0.02               & 0.00               & 1.00              & -0.02                 & 0.00                 & 1.22                 & -0.02                 & 0.00                 & 1.27                 & -0.02                 & 0.00                & 1.20                \\
Land area        & 0.44                 & 0.03                & 1.01               & 0.44                & 0.03               & 1.00              & 0.43                  & 0.03                 & 1.14                 & 0.44                  & 0.03                 & 1.19                 & 0.44                  & 0.03                & 1.08                \\
Floor area       & 0.53                 & 0.07                & 1.00               & 0.53                & 0.07               & 1.00              & 0.51                  & 0.07                 & 1.16                 & 0.53                  & 0.07                 & 1.20                 & 0.54                  & 0.07                & 1.13                \\
Age              & -0.03                & 0.00                & 1.01               & -0.03               & 0.00               & 1.00              & -0.03                 & 0.00                 & 1.21                 & -0.03                 & 0.00                 & 1.25                 & -0.03                 & 0.00                & 1.43                \\
Breadth          & 0.03                 & 0.01                & 1.01               & 0.03                & 0.01               & 1.00              & 0.03                  & 0.01                 & 1.19                 & 0.03                  & 0.01                 & 1.26                 & 0.03                  & 0.01                & 1.14                \\
Coverage ratio   & 0.80                 & 0.30                & 1.01               & 0.80                & 0.31               & 1.00              & 0.89                  & 0.29                 & 1.13                 & 0.87                  & 0.28                 & 1.21                 & 0.93                  & 0.27                & 1.30                \\
Floor area ratio & -0.12                & 0.05                & 1.01               & -0.12               & 0.05               & 1.00              & -0.09                 & 0.05                 & 1.15                 & -0.09                 & 0.05                 & 1.22                 & -0.11                 & 0.05                & 1.19               \\
\bottomrule
\end{tabular}
\end{table}

\section{Conclusion and Discussion}
\label{sec:end}

Integrated IPD–AD meta-analysis provides a powerful framework for synthesizing evidence across studies, but its efficiency depends critically on how AD is summarized and incorporated.
Within a unified CMLE framework, we compared several commonly available forms of AD, including marginal means of outcomes or covariates, covariate-stratified outcome summaries, and outcome-stratified covariate summaries, both when the IPD and AD arise from the same distribution and when they differ under dataset shift.
Our results show that outcome-stratified covariate summaries can yield substantial efficiency gains relative to the other two classes, particularly when the outcome is continuous.
In addition, we showed how CMLE can be extended to handle heterogeneity between IPD and AD through a general dataset shift model and developed a fast, non-iterative algorithm that improves numerical stability and scalability in high-dimensional applications with rich AD and many constraints.

These findings have important implications for the design, reporting, and curation of clinical and epidemiologic studies.
Outcome-stratified covariate summaries are routinely reported for binary outcomes in case–control designs but are far less common for continuous endpoints.
Our results suggest that providing such summaries for continuous outcomes can substantially improve the efficiency of downstream IPD–AD analyses.
We therefore recommend that trial reports and public data repositories routinely include outcome-stratified covariate summaries for continuous endpoints to support more informative and efficient evidence synthesis.

This paper has focused on parametric models.
An important direction for future research is to extend the constrained likelihood approach to more general semiparametric settings, such as the single-index model of \cite{DelecroixHH2003Efficient} and the effective dimension reduction model of \cite{Li1991Sliced}.
These formulations involve functional nuisance components in addition to finite-dimensional parameters, which raises new challenges for both estimation and computation.
These challenges warrant further development of constrained likelihood methods for semiparametric models, in settings with and without dataset shift, together with a rigorous investigation of their theoretical and numerical properties.

\bibliographystyle{apalike}
\bibliography{references}

\end{document}